%% file: main.tex
\documentclass[conference]{IEEEtran}
\usepackage{paralist}
\usepackage{multirow}
\usepackage[bookmarks=false]{hyperref}

\usepackage{cite}

\ifCLASSINFOpdf
\else
\fi
\usepackage{url}

\usepackage{tikz}
 \usetikzlibrary{arrows,positioning,shapes,shadows,decorations.shapes,automata,calc}

\tikzset{
    >=stealth',
    broker/.style={
           rectangle,
           rounded corners,
           draw=black, very thick,
           text width=6.5em,
           minimum height=2em,
           text centered},
    non_client/.style={
           ellipse,
           draw=black, very thick,
           text width=7.5em,
           minimum height=2em,
           text centered},       
    client/.style={
	   rectangle,
           draw=black, very thick,
           text width=5em,
           minimum height=2em,
           text centered},
    pil/.style={
           ->,
           thick,
           shorten <=2pt,
           shorten >=2pt,}
}
\tikzset{every picture/.style={font issue=\scriptsize},
         font issue/.style={execute at begin picture={#1\selectfont}}
        }
\tikzset{multiple/.style = {double copy shadow={shadow xshift=0.3ex,shadow
         yshift=-0.4ex,draw=black!30},fill=white,draw=black,thick,minimum height = 0.5cm,minimum
           width=2cm},
         ordinary/.style = {rectangle,draw,thick,minimum height = 0.5cm,minimum width=2cm}}

\usepackage[acronym]{glossaries}
\newacronym[plural=DFA,firstplural=deterministic finite automata (DFA)]{DFA}{DFA}{deterministic finite automaton}
\newacronym[plural=SULs,firstplural=systems under learning (SULs)]{SUL}{SUL}{system under learning}
\newacronym{MQTT}{MQTT}{Message Queuing Telemetry Transport}
\newacronym{IoT}{IoT}{Internet of Things}
\newacronym{TLS}{TLS}{transport layer security}
\newacronym{TCP}{TCP}{transmission control protocol}
\newacronym{UDP}{UDP}{user datagram protocol}
\newacronym{MAT}{MAT}{minimally adequate teacher}
\newacronym{QoS}{QoS}{quality of service}
\newacronym{EFSM}{EFSM}{extended finite state machine}
\newacronym{LTS}{LTS}{labelled transition system}
\newacronym{IOTS}{IOTS}{input-output transition system}
\newacronym{DHC}{DHC}{Direct Hypothesis Construction}
\newacronym{SMT}{SMT}{satisfiability modulo theories}
\newacronym{ISO}{ISO}{International Organization for Standardization}
\newacronym{ioco}{\textbf{ioco}}{input-output conformance}

\makeglossaries

\usepackage{amsthm}
\theoremstyle{definition}
\newtheorem{definition}{Definition}[section] 
\newtheorem{example}{Example}[section]
\theoremstyle{remark}

\usepackage{algpseudocode}
\usepackage{algorithm}
\algrenewcommand\Return{\State \algorithmicreturn{} }%

\newcommand{\var}[1]{\mathit{#1}}
\newcommand{\specialcell}[2][c]{%
  \begin{tabular}[#1]{@{}c@{}}#2\end{tabular}}

\makeatletter
\def\ps@IEEEtitlepagestyle{
  \def\@oddfoot{\mycopyrightnotice}
  \def\@evenfoot{}
}
\def\mycopyrightnotice{
  {\footnotesize
  \begin{minipage}{\textwidth}
  \centering
  \copyright 2017 IEEE.  Personal use of this material is permitted.  Permission from IEEE must be obtained for all other uses, in any current or future media, including reprinting/republishing this material for advertising or promotional purposes, creating new collective works, for resale or redistribution to servers or lists, or reuse of any copyrighted component of this work in other works.
  \end{minipage}
  }
}
\makeatother
\begin{document}

\title{Model-Based Testing IoT Communication via Active Automata Learning}

\author{\IEEEauthorblockN{Martin Tappler \quad  Bernhard K. Aichernig}
\IEEEauthorblockA{Institute of Software Technology \\
Graz University of Technology, Austria\\
\{martin.tappler, aichernig\}@ist.tugraz.at
}

\and
\IEEEauthorblockN{Roderick Bloem}
\IEEEauthorblockA{Institute of Applied Information Processing and Communications \\
Graz University of Technology, Austria\\
roderick.bloem@iaik.tugraz.at
}
}


\maketitle

\begin{abstract}
This paper presents a learning-based approach to detecting failures in reactive systems.
The technique is based on inferring models of multiple implementations of a common
specification which are pair-wise cross-checked for equivalence. Any counterexample
to equivalence is flagged as suspicious and has to be analysed manually. Hence, it is 
possible to find possible failures in a semi-automatic way without prior modelling. 

We show that the approach is effective by means of a case study. For this case study, 
we carried out experiments in which we learned models of five implementations
of MQTT brokers/servers, a protocol used in the Internet of Things. 
Examining these models, we found several violations of the MQTT specification. 
All but one of the considered implementations showed faulty behaviour.
In the analysis, we discuss effectiveness and also issues we faced.

\end{abstract}

%

\IEEEpeerreviewmaketitle

\input{intro}
\input{related_work}

\input{prelim}
\input{approach}
\input{case_study}

\input{conclusion}


\section*{Acknowledgment}

This work was supported by the TU Graz LEAD project "Dependable Internet of Things in Adverse Environments".
The authors would like to thank the LEAD project members 
Masoud Ebrahimi, Franz Pernkopf, Franz R{\"o}ck, and Tobias Schrank for fruitful discussions, 
and Florian Lorber, Richard Schumi and the anonymous reviewers
for their feedback. 
Additionally, we would like to thank the developers of LearnLib.

\vspace{1.2cm}

\IEEEtriggeratref{36}

\bibliographystyle{IEEEtranS}
\bibliography{biblio}

\vfill

\end{document}

%% file: intro.tex
\section{Introduction}

Active automata learning has gained increasing attention of the verification and testing community 
in recent years. There exist several different approaches to this kind of learning. Many of 
them are based on or related to the $L^*$ algorithm by Angluin \cite{Angluin1987}. As such these approaches
share a strong connection to conformance testing~\cite{Berg2005}. In both areas, learning and conformance testing,
the goal is to gain knowledge about the behaviour of a black-box system, by executing tests/queries\footnote{Tests 
are often called (membership/output) queries in active automata learning.} 
and analysing corresponding observations. However, in the former we are interested in the synthesis aspect, i.e. we 
want to infer a model, whereas in the latter, we perform an analysis task, i.e. we check conformance to a given 
model.

This opens up the possibility to combine these approaches. Aarts et al.~\cite{Aarts2012a} have for instance
shown how to combine learning, testing and verification. They learned the model of a reference implementation
of the bounded retransmission protocol and checked equivalence between this model and several mutated (faulty) implementations via
two different techniques. (1) They performed model-based testing of the mutants using the learned reference model. 
(2) Additionally, they also learned models of the mutants and 
subsequently checked equivalence between the inferred models of the specification and of each of the mutants. 
In this paper, we will follow an approach similar to the latter. 
Both approaches differ most significantly in the kind of implementations considered.
They actually generated Java applications from models with a known structure. Furthermore, 
the faulty implementations have been created artificially by seeding known errors into the reference model. 
On the contrary, we do not know anything about the structure of the analysed implementations.
We merely know that they implement a common specification given in natural language.

\begin{figure}[t]
 \centering
 \begin{tikzpicture}[node distance = 0.5cm]
   \node [multiple] (impl) {\scriptsize Implementations};
   \node [multiple, below = of impl] (abs_model) {\scriptsize Abstract Models};
   \node [ordinary, below left = 0.2cm and 0cm of abs_model] (abs_model1) {\scriptsize Single Abstract Model};
   \node [ordinary, below right = 0.2cm and 0cm  of abs_model] (abs_model2) {\scriptsize Single Abstract Model};
   \node[multiple, below = 1.3cm of abs_model] (differences) {\scriptsize Differences};
   \node [ordinary, right = of differences] (standard) {\scriptsize Standards Document};
   \node [multiple, below = of differences] (bugs) {\scriptsize Bugs};
   
   \draw[->,thick] (impl) -- node[below right  = -0.1cm and 0cm]{\scriptsize Learning} (abs_model);
   \draw[->,thick] (abs_model) |- node[below = 0cm ]{\scriptsize Choose Pair} (abs_model1);
   \draw[->,thick] (abs_model) |-  (abs_model2);
   \draw[->,thick] (abs_model1.south) --+(0,-0.2cm) -| node[below right = 0 cm and 0 cm]{\scriptsize Check Equivalence} (differences);
   \draw[->,thick] (abs_model2.south) --+(0,-0.2cm) -|  (differences);
   \draw[->,thick] (differences) -- node[below right = -0.1 cm and 1 cm]{\scriptsize Analyse Manually} (bugs);
   \draw[->,thick] (standard.south) --+(0,-0.2cm) -|  (bugs); 
   
 \end{tikzpicture}
\caption{Overview of bug-detection process.}
\label{fig:process_overview}
\end{figure}
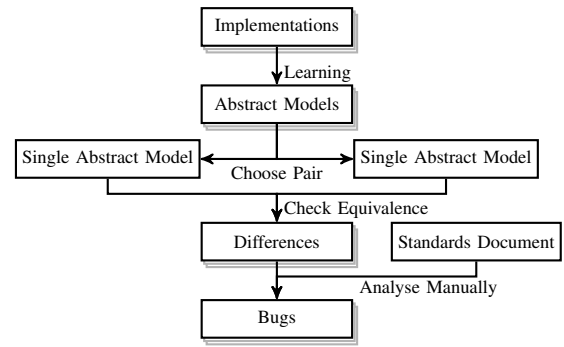

In order to detect faults in the considered implementations, we thus propose the following learning-based approach. In a first phase, we learn 
(Mealy-machine) models of several 
different implementations of some standardised protocol or operation. These models are then pair-wise cross-checked with 
respect to some conformance relation. All counterexamples to conformance are then analysed manually by consulting 
a given standards document. This may either reveal a bug in one or both implementations corresponding to the counterexample, 
or it may reveal an underspecification of some aspect of the standard. 
The process we follow is also depicted in \figurename~\ref{fig:process_overview}.

This approach apparently cannot detect all possible faults because specific faults may be 
implemented by all examined implementations. In addition to that, the fault-detection capabilities
are limited by the level of abstraction used for learning. 
This gives rise to two research questions we aim at answering in the following.
Is it possible to effectively detect non-trivial faults using our approach despite the necessary
severe abstraction? What are the limitations of our approach? Related to 
the second question, we will discuss opportunities for future research
in order to mitigate the identified limitations. 

For this purpose we will analyse the behaviour of five different brokers implementing
\gls*{MQTT} version 3.1.1~\cite{MQTT_spec}, a protocol standardised
by the \gls*{ISO}~\cite{ISO_MQTT}. The \gls*{MQTT} protocol is a lightweight
publish/subscribe protocol and therefore well-suited for resource-constrained environments
such as the \gls*{IoT}. This is the main reason we have chosen \gls*{MQTT} for our case study
as we want to investigate verification techniques in the context of the \gls*{IoT}. 
Furthermore, we consider broker implementations since they constitute central communication
units, hence it is essential to assure their correct operation for a reliable communication
in the \gls*{IoT}. Basically, brokers allow clients to subscribe and to publish to topics. If 
a message is published, the broker forwards it to all clients with matching subscriptions. 

The main contribution of this paper is thus the presentation and empirical evaluation of the mentioned approach
based on learning experiments with five different black-box systems. 
More concretely, we learned models of five different \gls*{MQTT} brokers, systems used in
\gls*{IoT} communication. In our analysis, we will 
discuss failure-detection with a focus on required effort, issues
related to runtime, and general challenges.

We are not aware of any case study applying conformance checking between learned models
in a purely black-box setting. 
Neither do we know of any case study focusing on the verification of
implementations of the \gls*{IoT} protocol \gls*{MQTT}. 

The structure of this paper is as follows. In Section \ref{sec:rel_work}, we will discuss related
work in the area of testing and verification. In Section \ref{sec:prelim}, we will introduce the used modelling
formalism and active automata learning. 
Section \ref{sec:approach} introduces the approach we follow in our case study. 
We will present implementation details and results obtained from learning experiments
in Section \ref{sec:case_study}.
We conclude in Section \ref{sec:discussion} with a summary of our findings and a discussion 
of future work. 

%% file: related_work.tex
\section{Related Work}
\label{sec:rel_work}

In this section, we will review related work with a focus on the combination of 
active automata learning and verification. We already mentioned the work by Aarts et al.~\cite{Aarts2012a}
in the introduction. Although they also used Mealy machines and a similar technique to distinguish mutated (faulty) implementations 
from a reference implementation, our setting is different. While Aarts et al.\ consider learning of
more complex models in terms of states and number of inputs, we deal with different challenges.
The challenges we face are mainly related to the network communication. We have to deal with long 
response times and with completely unknown system structure. 
Basically, we do not know whether our implementations behave like
Mealy machines, a property fulfilled by their implementations generated from a known UPPAAL model.
Additionally, they created the mutated implementations by seeding faults into the reference implementation, thus 
they also know beforehand that the checked systems share a similar structure.

We apply both testing and exhaustive equivalence checking in our case study. The former is used 
during learning and the latter during the analysis of learned models. Early work in this area has been performed
by Peled et al.~\cite{Peled2002} and Groce et al.~\cite{Groce2002}. They combine testing and model checking, 
in order to analyse black-box systems as well. Additionally, they also apply active automata learning to infer 
the structure of the analysed systems from the observations made during testing. 

We make use of the assumption that a learned model faithfully represents the corresponding black-box system. Hence,
we can simulate tests of a system by executing tests on the learned automaton. This is possible since we already
tested for equivalence/conformance between the black-box system and the hypothesis automaton during learning. 
Berg et al.~\cite{Berg2005} discuss the correspondence between conformance testing and learning in more detail 
and show similarities and differences. 

In the analysis of the \gls*{TLS} protocol, de Ruiter and Rutten consider a similar learning setup~\cite{de_ruiter2015}.
They investigate the behaviour of several implementations of the \gls*{TLS} protocol via active automata learning, but they manually 
analyse the inferred models. While we are interested in any error that can be found on state-machine level, 
they specifically target security-related flaws. Beurdouche et al.~\cite{Beurdouche2015}
also targeted state-machine based flaws. They followed a test-based approach, but generated 
tests from a known model via some heuristics. Thereby they checked for specific faults, like faults 
related to skipping mandatory steps in a protocol. 

Fiter{\u{a}}u-Bro{\c{s}}tean et al.\ also performed case studies involving active learning of models of several implementations of 
a single protocol~\cite{Fiterau_Brostean2014,Fiterau_Brostean2016}. More concretely, they learned Mealy-machine models
of \gls*{TCP} implementations. They additionally applied model-checking~\cite{Fiterau_Brostean2016} in order to 
verify properties of the composition of client and server implementations.

%% file: prelim.tex
\section{Preliminaries}

\label{sec:prelim}

\subsection{Mealy Machines}

We will use Mealy machines as modelling formalism because they are well-suited to model 
reactive systems and they have successfully been used in contexts combining learning
and some form of verification~\cite{Margaria2004,de_ruiter2015,Fiterau_Brostean2016}. In addition to that, 
the application of Mealy machines allows us to use the existing Java-library LearnLib~\cite{isberner2015}
which provides efficient algorithms for learning Mealy machines. 

Basically, Mealy machines are finite state automata with inputs and outputs. The execution 
of such a Mealy machine starts in an initial state and by executing inputs it changes its state.
In addition to that, exactly one output is produced in response to each input. 
We will refer to Mealy machines also as state machines in the remainder of this paper.
Formally, Mealy machines can be defined as follows. 

\begin{definition}[Mealy Machines]
 A Mealy machine is a $6$-tuple $\langle Q, q_0, I,O,  \delta, \lambda\rangle$ where 
 \begin{itemize}
  \item $Q$ is a finite set of states,
  \item $q_0$ is the initial state, 
  \item $I$/$O$ is a finite set of input/outputs symbols,
  \item $\delta : Q \times I \rightarrow Q$ is the state transition function, and
  \item $\lambda : Q \times I \rightarrow O$ is the output function
 \end{itemize}
\end{definition}
\ifdefined\FULL
Both $\delta$ and $\lambda$ define reactions of the system in given states in response to inputs so we usually combine these functions,
forming the transition relation $\rightarrow\ \subseteq Q \times I \times O \times Q$. An element $(q,i,o,q')$, also called transition,
is contained in the transition relation $\rightarrow$ iff $\delta(q,i) = q'$ and $\lambda(q,i) = o$.
\fi

We require Mealy machines to be input enabled and deterministic. 
The former demands that an output and a successor state must be defined for all inputs and all states, i.e. 
$\delta$ and $\lambda$ must be surjective. A Mealy machine is deterministic if it defines at most one 
output and successor state for every pair of input and state, thus $\delta$ and $\lambda$ must be  
functions in the mathematical sense.

We will now introduce some notational conventions for sequences of input/output symbols $s \in S^*$, where $S = I$ or $S=O$. 
Let $s' \in S^*$ be another sequence, then $s \cdot s'$
denotes the concatenation of these sequences. 
The empty sequence is represented by $\epsilon$.
We implicitly lift single elements to sequences, thus for $e \in S$ we have 
$e \in S^*$. As a result, the concatenation $s \cdot e$ is also defined.

Furthermore, $\delta$ and $\lambda$ are extended to sequences of inputs in the standard way. 
Let $s \in I^*$ be a sequence of inputs and $q\in Q$ be a state
of a Mealy machine, then $\delta(q,s) = q' \in Q$ is the state reached by executing $s$ starting in state $q$.
Given a sequence of inputs $s \in I^*$ and a state $q\in Q$, the output function $\lambda(q,s) = t \in O^*$ returns 
the outputs produced in response to $s$ executed in state $q$. 

Finally we need a basis for determining whether two Mealy machines are equivalent. Equivalence
is usually defined with respect to outputs~\cite{Aarts2012}, i.e.
two deterministic Mealy machines are equivalent if they produce the same outputs for all input sequences.
We say that a Mealy machine $\langle Q_1, {q_0}_1, I,O,  \delta_1, \lambda_1\rangle$
is equivalent to another Mealy machine $\langle Q_2, {q_0}_2, I,O,  \delta_2, \lambda_2\rangle$ iff
$\forall s \in I^* : \lambda_1({q_0}_1,s) = \lambda_2({q_0}_2,s)$. A counterexample to equivalence is thus an $s\in I^*$
such that $\lambda_1({q_0}_1,s) \neq \lambda_2({q_0}_2,s)$.

\ifdefined\FULL
\begin{example}[Mealy Machine with Two Inputs]
\label{ex:mealy_machine}
 The following example shows a simple Mealy machine with two inputs and three Outputs. Basically, the system 
 represented by the Mealy machine can acknowledge connections, answer ping messages and disconnect a client
 which tries to connect twice. 
 It actually shows a part of a learned model of an \gls*{MQTT} broker.
 
 The Mealy machine is defined by $\langle Q, q_0, I,O,  \delta, \lambda\rangle$ where 
 \begin{compactitem}
  \item $Q = \{q_0,q_1\}$
  \item $I = \{\mathit{Connect},\mathit{Ping}\}$
  \item $O = \{\mathit{ConnAck},\mathit{Pong}, \mathit{ConnectionClosed}\}$
  \item $\delta$ is defined by:
  $\begin{aligned}
         \delta(q_0,\mathit{Connect}) &= q_1 & \delta(q_0,\mathit{Ping})&= q_0\\
         \delta(q_1,\mathit{Connect}) &= q_0 & \delta(q_1,\mathit{Ping})&= q_1\\
        \end{aligned} 
        $
  \item $\lambda$ is defined by: $\begin{aligned}
         \lambda(q_0,\mathit{Connect}) &= \mathit{ConnAck} \\ \lambda(q_0,\mathit{Ping})&= \mathit{ConnectionClosed}\\
         \lambda(q_1,\mathit{Connect}) &= \mathit{ConnectionClosed} \\ \lambda(q_1,\mathit{Ping})&= \mathit{Pong}\\
	\end{aligned}$
 \end{compactitem}
The graphical representation of $\mathcal{M}$ is shown in \figurename \ref{fig:example_mealy_m}.
\begin{figure}
\centering
 \begin{tikzpicture}[>=stealth',shorten >=1pt,auto,node distance = 3.5cm]
  \node[initial,state] (q_0)      {$q_0$};
  \node[state] (q_1) [right of = q_0] {$q_1$};
  
    \path[->] (q_0)  edge [loop above,min distance=15mm,in=70,out=110] node[text width = 14em,text centered] {$\mathit{Ping} /$ $\mathit{ConnectionClosed}$} (q_0)
  edge [bend left] node[above] {$\mathit{Connect}/ \mathit{ConnAck}$} (q_1)
  (q_1)  edge [loop above,min distance=15mm,in=70,out=110] node {$\mathit{Ping} / \mathit{Pong}$} (q_1)
  edge [bend left] node[below] {$\mathit{Connect}/ \mathit{ConnectionClosed}$} (q_0);
 \end{tikzpicture}
\caption{A simple Mealy machine acknowledging connections and responding to ping messages}
\label{fig:example_mealy_m}
\end{figure}

\end{example}
\fi

\subsection{Active Automata Learning}

In the following, we will consider learning algorithms operating in 
the \gls*{MAT} framework proposed by Angluin~\cite{Angluin1987}. These algorithms
infer models of black-box systems, also referred to as \glspl*{SUL}, through interaction with a 
so-called teacher. 

\subsubsection{Minimally Adequate Teacher Framework}
The interaction is carried out via  two types of queries posed by the learning algorithm and 
answered by a \acrlong*{MAT}. 
These two types of queries are usually called \emph{membership queries} and \emph{equivalence queries}. 
In order to understand these basic notions of queries consider that
Angluin's original $L^*$ algorithm was used to learn a \gls*{DFA} representing 
a regular language known to the teacher~\cite{Angluin1987}. Given some alphabet, the $L^*$ algorithm 
repeatedly selects strings  and asks membership queries in order to
check whether these strings are in the language to be learned. The teacher may answer either
\emph{yes} or \emph{no}. 

After some queries the learning algorithm uses the knowledge gained so far and 
forms a hypothesis, i.e. a \gls*{DFA} consistent with the obtained information which
should represent the regular language under 
consideration. The algorithm presents the hypothesis to the teacher and issues an equivalence
query in order to check whether the language to be learned is equivalent to the language
represented by the hypothesis automaton. The response to this kind of query is either
\emph{yes} signalling that the correct \gls*{DFA} has been learned or a counterexample to equivalence.
Such a counterexample is a witness showing that the learned model is not yet correct,
i.e. it is a word from the symmetric difference of the language under learning and 
the language accepted by the hypothesis. 

If a counterexample is provided then learning algorithms incorporate this counterexample
into their data structures and start a new \emph{round} of learning. The new round 
again involves membership queries and a concluding equivalence query. 

This general mode of operation is used by basically all algorithms in 
the \gls*{MAT} framework with some adaptations. These adaptations may for instance enable the
learning of Mealy machines which we will describe in the following.

\subsubsection{Learning Mealy Machines}


Margaria et al.~\cite{Margaria2004} and Niese~\cite{Niese2003} were
one of the first to infer Mealy-machine models of reactive systems 
by applying a learning algorithm based on $L^*$. 
Another learning algorithm for Mealy machines, also based on $L^*$, has been presented by Shahbaz and Groz~\cite{shahbaz_groz_2009}.
They basically reuse  the structure of $L^*$, but instead of membership queries, they pose \emph{output queries}. Thus,
instead of checking whether a string is in some language, they provide an input string to the teacher and the teacher responds
with the corresponding output string. 

For a more practical discussion of learning consider the instantiation of a teacher. Usually we 
want to learn the behaviour of a black-box \gls*{SUL} of which we only know the input and output interface. 
Hence, output queries are conceptually simple: inputs are provided to the \gls*{SUL} and it produces
some outputs. However, there is a slight difficulty hidden. Like Angluin~\cite{Angluin1987}, 
Shahbaz and Groz~\cite{shahbaz_groz_2009} assume that output queries provide outputs in response
to inputs executed from the \textbf{initial} state. Consequently, we need to have some means to reset a
system. Since we are dealing with a black-box system, we normally cannot check for equivalence with a hypothesis. 
In practice, it is thus necessary for a teacher used in a learning algorithm to approximate equivalence queries somehow. 
This can for instance be achieved via conformance testing as implemented in LearnLib~\cite{isberner2015}.

\begin{figure}
\centering
 \begin{tikzpicture}[thick]
  \node[draw,text depth = 1cm,minimum width=4cm, minimum height = 5cm] (teacher){};
  \node[font=\bf] at ([yshift=-1em]teacher.north) {\scriptsize Teacher};
\node[draw, font=\bf,text width = 5em, text centered, minimum height = 1.5 cm] at ([xshift=-1.0cm,yshift=-1.25cm]teacher.north)(mbt){\scriptsize Model-Based Testing Tool};
\node[draw, font=\bf,text width = 5em, text centered, minimum height = 1.5 cm] at ([xshift=-1.0cm,yshift=1.25cm]teacher.south)(sul){\scriptsize \gls*{SUL}};

\node[draw,font=\bf, text depth = 1cm,minimum width=1.75cm, text width=1.55cm, text centered, minimum height = 5cm, right = 2.5 cm of teacher] (learner){\scriptsize Learning Algorithm};
\draw[->] (teacher.33 -|learner.west) -- node[above right = 0cm and -1.35cm, text width = 10em,align=right]{\scriptsize Equivalence Query} (teacher.33);
\draw[->] (teacher.17) -- node[above right = 0cm and -1.35cm, text width = 10em,align=right]{\scriptsize Yes / Counterexample}  (teacher.17 -|learner.west);
\draw[->] (teacher.33) -- node[above right = 0cm and -1.05cm, text width = 8em,align=right]{\scriptsize Perform Tests}  (teacher.33 -|mbt.east);
\draw[->] (teacher.17 -|mbt.east) -- node[above right = 0cm and -1.05cm, text width = 8em,align=right]{\scriptsize All Pass / Failed Test}  (teacher.17);

\draw[->] (teacher.343 -|learner.west) -- node[above right = 0cm and -1.35cm, text width = 10em,align=right]{\scriptsize Output/Membership Query} (teacher.343);
\draw[->] (teacher.327) -- node[above right = 0cm and -1.35cm, text width = 10em,align=right]{\scriptsize Query Output}  (teacher.327 -|learner.west);
\draw[->] (teacher.343 ) -- node[above right = 0cm and -1.05cm, text width = 8em,align=right]{\scriptsize Reset + Inputs} (teacher.343-|sul.east);
\draw[->] (teacher.327-|sul.east) -- node[above right = 0cm and -0.8cm, text width = 7em,align=right]{\scriptsize Outputs} (teacher.327);

\draw[<-] (mbt.245) -- node[left = -0.3 cm] {\scriptsize Outputs} (sul.north -| mbt.245);
\draw[->] (mbt.295) -- node[right] {\scriptsize Reset + Inputs} (sul.north -| mbt.295);

 \end{tikzpicture}
\caption{The interaction between \gls*{SUL}, teacher and learning algorithm (based on a figure by Smeenk et al.~\cite{Smeenk2015}).}
\label{fig:interaction_teacher_learner}
\end{figure}
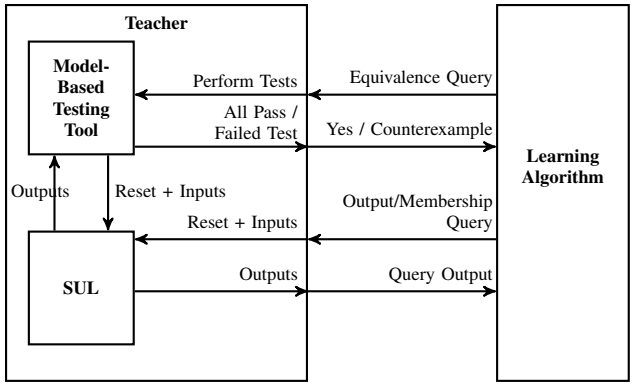

To summarise, a learning algorithm for Mealy machines relies on three operations:
\begin{compactdesc}
 \item[reset:] resets the \gls*{SUL}
 \item[output query:] performs a single test executing a sequence of inputs and recording the outputs
 \item[equivalence query:] performs a set of tests comparing the outputs of the \gls*{SUL} and the current hypothesis
\end{compactdesc}
Hence, the teacher is usually some component interacting with the \gls*{SUL} in order to implement these operations. 
An equivalence query results in a positive answer if all tests pass, i.e. the \gls*{SUL} produces the same 
outputs as the hypothesis automaton. If there is a test for which their outputs differ, the corresponding sequence
of inputs is presented to the learning algorithm as a counterexample. The interaction between \gls*{SUL},
teacher and learning algorithm is also depicted in \figurename{}~\ref{fig:interaction_teacher_learner}.

Note that due to the incompleteness of testing the learned model may be incorrect if the equivalence query is replaced with conformance testing. 
If for instance the W-method by Vasilevskii~\cite{Vasilevskii1973} and Chow~\cite{Chow1978}
is used to derive a conformance test suite, the learned model 
may not be correct if assumptions placed on the maximum number of states of the \gls*{SUL} do not hold.

%% file: approach.tex
\section{Approach}
\label{sec:approach}

In this section, we will discuss our approach in more detail.
We will start with a discussion of learning Mealy-machine models in which we 
also highlight some technical considerations to enable the task of learning such models.

\subsection{Learning}

\ifdefined\FULL
Although Mealy machines are well-suited to model reactive systems, \gls*{MQTT}-broker implementations 
may not behave exactly like Mealy machines. This results from the fact that (1) the actual durations of delays between messages may be 
of relevance which are usually abstracted away and (2) some implementations do not behave deterministically. Both issues are
actually related, as varying network latency may cause the arrival of messages to appear non-deterministically. 
Additionally, delays may also vary in dependence of incompletely known influences.

Since complex reactive systems operating in a networked environment
are likely to show a certain degree of non-determinism, approaches
to learning non-deterministic automata have been proposed~\cite{Khalili_Tacchella2014,Volpato_Tretmans2015}. 
However, \gls*{MQTT}-brokers behave largely deterministic under certain restrictions. For this reason and because the active 
automata learning library LearnLib~\cite{isberner2015} provides mature and efficient implementations of learning algorithms for deterministic
Mealy machines, we have chosen to restrict the work in this case study to deterministic systems. Nevertheless, we will 
discuss issues related to non-determinism and point out directions for future work in relevant sections.
\fi

\subsubsection{Architecture}

Two aspects influence the architecture of a learning environment for \gls*{MQTT}. Firstly, we need to account for dependencies between
clients. Unlike in pure client/server-settings, like in the \gls*{TLS} protocol~\cite{de_ruiter2015} or the 
\gls*{TCP}~\cite{Aarts2010,Fiterau_Brostean2016}, it is not sufficient to simulate one client to 
adequately infer a model of the server/broker in \gls*{MQTT}. We need to control multiple clients and record
the messages each one has received.

Secondly, we need to cope with the enormous amount of possible inputs, i.e. the large number of packets we can send to the brokers. This, however, 
is a general problem of active automata learning in the \gls*{MAT} framework and an issue for learning almost all non-trivial
systems. To deal with this problem, we introduce a mapper component performing abstraction and concretisation~\cite{Howar2010,Aarts2010,Aarts2012}.
However, unlike in the cited work, we do not refine our abstractions in an iterative manner, but rather use a static mapper throughout the learning
phase. If non-determinism arising from abstraction or from the processing of outputs in general was detected, we manually adapted
the learning setup in an appropriate way. 

\figurename{} \ref{fig:learning_architecture} shows the architecture of our learning setup. 
The \gls*{SUL}, an \gls*{MQTT} broker, is shown on the left-hand side. In order to learn its behaviour we control 
its environment made up of several clients which basically consist of the adapter blocks and the client-interface blocks.
The adapter handles communication-related tasks, whereas the client-interface components implement a
client library with a simple interface and default values for control-packet parameters. 

The right-hand side of \figurename{} \ref{fig:learning_architecture} shows the components responsible for learning.
LearnLib implements several algorithms for actively learning Mealy machines. During the execution of such an
algorithm, LearnLib interacts with a mapper component by choosing and executing
one of the available abstract inputs. The mapper concretises abstract inputs by choosing one of the clients
and executing some action with concrete values. Note that the sending of packets and serialisation tasks performed during 
an action are handled by the clients. As soon as outputs are available, the mapper collects them from all clients and abstracts
them, creating one abstract output symbol in response to each abstract input. 

\tikzset{
    block/.style={
           rectangle,
           rounded corners,
           draw=black, thick,
           text width=5em,
           minimum height=2em,
           text centered}
           }
\tikzset{decorate sep/.style 2 args=
{decorate,decoration={shape backgrounds,shape=circle,shape size=#1,shape sep=#2}}}

\begin{figure}
 \centering
 \begin{tikzpicture}[node distance = 0.5cm]
  
 \node[block] (broker) {\scriptsize \gls*{MQTT} Broker};
 
 \node[block, above right = 1 cm and 0.25 cm of broker] (adapter1) {\scriptsize Adapter $1$};
 \node[block, below = 0.3cm of adapter1] (adapter2) {\scriptsize Adapter $2$};
 \node[block, below right = 0.8 cm and 0.25 cm of broker] (adapter3) {\scriptsize Adapter $n$};
 \draw[decorate sep={0.4mm}{2.3 mm}, fill] (adapter2) -- (adapter3);
 
  \draw[<->,thick] (broker) |- node(description_broker_adapter_pos){} (adapter1);
  \draw[<->,thick] (broker) |- (adapter2);
  \draw[<->,thick] (broker) |- (adapter3);

 \node[block,right = of adapter1] (client1) {\scriptsize client interface $1$};
 \node[block,right = of adapter2] (client2) {\scriptsize client interface $2$};
 \node[block, right = of adapter3] (client3) {\scriptsize client interface $n$};
 \draw[decorate sep={0.4mm}{2.3 mm}, fill] (client2) -- (client3);

  \draw[<->,thick] (adapter1) edge node(description_adapter_client_pos){}  (client1);
  \draw[<->,thick] (adapter2) edge (client2);
  \draw[<->,thick] (adapter3) edge (client3);
 
 \node[minimum width=6em] at (broker -| client1)(dummy_mapper_positioner){};
 \node[block, right = 0.25 cm of dummy_mapper_positioner] (mapper) {\scriptsize Mapper};
 
  \draw[<->,thick] (client1) -|node(description_client_mapper_pos){} (mapper);
  \draw[<->,thick] (client2) -| (mapper);
  \draw[<->,thick] (client3) -|  (mapper);
  
 \node[block, right = 0.55 cm of mapper] (learnlib) {\scriptsize LearnLib};
 
  \draw[<->,thick] (mapper) edge node(description_mapper_learnlib_pos){} (learnlib);
 
 
 \node[above = 0.9cm of description_mapper_learnlib_pos, text width = 4.5em, align = flush right] (description_mapper_learnlib){\scriptsize $\triangleleft$ abstract \\ inputs};
 \node[below = 0cm of description_mapper_learnlib, text width = 4.5em, align = flush right] {\scriptsize $\triangleright$ abstract \\ outputs};
 \node[above left = 0.9cm and 0.2cm of description_client_mapper_pos, text width = 4.5em, align = flush right] (description_client_mapper){\scriptsize $\triangleleft$ concrete \\ inputs};
 \node[below = 0cm of description_client_mapper, text width = 4.5em, align = flush right] {\scriptsize $\triangleright$ concrete \\ outputs};
 
 \node[above left = 0.9cm and -0.7cm of description_adapter_client_pos, text width = 4.5em, align = flush right] (description_adapter_client){\scriptsize $\triangleleft$ control \\ packets};
 \node[below = 0cm of description_adapter_client, text width = 4.5em, align = flush right] {\scriptsize $\triangleright$ control \\ packets};
 
  \node[above right = 0.9cm and -0.4cm of description_broker_adapter_pos, text width = 5.5em, align = flush right] (description_broker_adapter){\scriptsize $\triangleleft$ bytes via TCP/IP};
 \node[below = 0cm of description_broker_adapter, text width = 5.5em, align = flush right] {\scriptsize $\triangleright$ bytes via TCP/IP};

 \end{tikzpicture}
 \caption{The architecture we used for learning of Mealy-machine models of \gls*{MQTT} brokers.}
\label{fig:learning_architecture}
\end{figure}
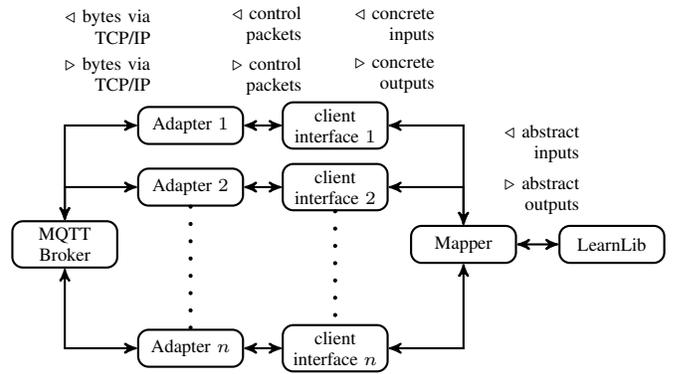

\subsubsection{Practical Considerations}
%

\paragraph{Timeouts}

We already noted that there are delays between the transmission of messages to the broker and the receipt of corresponding responses.
These are inevitably present since network communication is involved, which is actually asynchronous. In addition to that 
a system may not send any message, i.e. not produce any output at all.
A more faithful model of such a system would thus for instance be a timed automaton~\cite{AlurDill1994}. 

However, instead of changing the modelling
formalism we followed a more pragmatic approach. We basically set a configurable timeout for the receipt of messages 
like de Ruiter and Poll~\cite{de_ruiter2015}. All messages received before reaching this timeout are processed by 
the mapper component to form an abstract output symbol. If we do not receive any message,  we say that the system is
quiescent and the mapper produces the corresponding output symbol \emph{Empty}. 
This is similar to the way the absence of outputs is handled in 
\gls*{ioco} testing except that there quiescent behaviour is represented by the symbol $\delta$~\cite{Tretmans96}.

\paragraph{Processing Outputs}

There are several steps involved in the processing of messages received
from a broker. After deserialisation, relevant information is extracted
from messages. Since a single client may receive multiple messages in response 
to a single input sent by one of the clients, these message groups have to be
processed to create one output per client. 
We therefore sort the messages received by each client alphabetically to ensure
determinism. This is necessary as messages may arrive in any order. 
Afterwards we concatenate the sorted messages. 
If a client does not receive any message, we interpret this as having received
a single message \emph{Empty}. The outputs of individual clients are concatenated
to create a single abstract output.

While some messages may arrive in any order, the \gls*{MQTT} specification 
also places restrictions on message ordering~\cite{MQTT_spec}.
As a result, we lose relevant information by sorting.
Since alternatively we would need to learn non-deterministic models,
this way of handling outputs represents a tradeoff between completeness
and efficiency.

\paragraph{Restrictions Placed on \gls*{MQTT} Functionality}

We had to exclude some features of \gls*{MQTT} from our analysis as 
they cannot adequately be modelled using Mealy machines. The \gls*{MQTT} specification, e.g., 
includes a ping functionality which would require learning
of time-dependent behaviour. 

Additionally, there are dependencies between sent and received data, e.g.,
identifiers sent in acknowledgements should match identifiers of the acknowledged
messages. For this reason, we send acknowledgements automatically from client
to broker and do not learn behaviour related to that. 
This opens up a possibility
for future work: it would be possible to carry out experiments with the tool Tomte~\cite{Aarts2012,Aarts2015}. This tool
is able to learn a restricted class of \glspl*{EFSM}, Mealy machines with guards and state updates. These \glspl*{EFSM} 
are expressive enough to model the acknowledgement mechanism.

It should be emphasised that the excluded features do not affect the parts of \gls*{MQTT}
we learn. 

\subsection{Learning-Based Testing}

\label{sec:model_based_testing_process}

We describe our approach to testing in the following. 
Roughly speaking, we test conformance between implementations and flag traces leading
to conformance violations as suspicious. We do so by learning 
models of the concerned systems and subsequent equivalence checking of those models. 

For a more in-depth discussion of the topic, 
assume that a model learned with the approach described above faithfully represents the \gls*{SUL} under the 
abstraction applied by the mapper. As a result, we can execute tests on the model and thereby implicitly
perform tests on the \gls*{SUL} under the same abstraction. In other words, we can simulate a testing 
environment similar to \figurename{} \ref{fig:learning_architecture}, but with LearnLib replaced by 
a conformance testing component.

Note that this differs from the usual approach to model-based testing~\cite{utting2012}. 
In general, a model is assumed to be given which can be used for generating tests and as a test oracle.
The former is still possible with learned models, i.e. we can generate tests according to some criterion.
The latter, however, is problematic as we cannot be sure whether the model is correct.
A model learned by observing a faulty implementation will also be incorrect. 
To circumvent this problem, we use the learned model of another implementation as oracle. 

We thus test whether some learned model conforms to another learned model and thereby we implicitly test conformance
between the implementations. Since we do not know anything about the correctness of either implementation, 
we do not consider conformance violations to be errors, but rather flag traces leading to conformance violations
as suspicious. After performing all tests of a conformance test suite, we manually investigate all traces
which show non-conformance in order to determine whether any of the implementations violates the specification (i.e. the \gls*{MQTT} specification~\cite{MQTT_spec}).

This may reveal zero, one or two errors depending on whether the specification allows some
freedom of implementation for the corresponding functionality, and whether one or both implementations implemented
the functionality in a wrong way. 
Not all errors are detected for two reasons: (1) it may not be possible 
to detect some errors because of abstraction and (2) we do not detect an error if it is equally present in all implementations.
The first problem is inherent to model-based testing, while the second problem is directly related to our 
approach. To overcome this issue and to decrease the probability of missing errors we compared the behaviour
of five implementations instead of just two. 


In order to cope with the large number of abstract inputs, we decided to create several sets of abstract input alphabets.
In other words, we learned distinct models for subsets of the set of all abstract inputs. 
We thus had to implement as many mappers as input subalphabets. 
These subsets of inputs have been chosen in a way such that inputs
within some subset have interesting dependencies. As we thereby also place implicit assumptions about
independence relationships between inputs from different sets, the effectiveness of fault-detection may be impeded by 
this approach. Issues such as the effectiveness of the overall approach will be discussed 
in Section~\ref{sec:case_study}. 
A similar approach has been followed by Smeenk et al.~\cite{Smeenk2015} for 
equivalence testing. In addition to the complete alphabet containing
all inputs, they identified a subset of inputs relevant to initialisation which they tested more thoroughly. 

The separation into subsets of the complete input alphabet 
led to the following model-based testing process. \\
 For each input alphabet:
  \begin{compactenum}[1)]
  \item Learn a model $m_i$ of each implementation $i$
  \item For each pair $(m_i,m_j)$ of learned models:
  \begin{compactenum}[1)]
  \item Check equivalence
  \item For each counterexample $c$ to equivalence
  \begin{compactenum}[1)]
    \item Test $c$ on implementations $i$ and $j$
    \item Analyse manually if outputs of $i$ and $j$ are correct
  \end{compactenum}
  \end{compactenum}
  \end{compactenum}

Note that if we find a counterexample to equivalence, i.e. a suspicious trace $c$, we test it 
on the corresponding implementations. We do so to ensure that $c$ actually shows a difference 
between the implementations and is not the result from an error introduced by learning. 
Although active automata learning is in general sound, this may happen because we only approximate equivalence
queries by conformance testing.

%
As we check conformance on model-level, we can also use techniques 
other than testing. We could, e.g., use external tools such as CADP to check equivalence~\cite{aichernig_delgado_2006}, 
encode the problem as reachability problem and use \gls*{SMT} solvers for the task~\cite{Aichernig2015}, 
or use a graph-based approach~\cite{Brandl2010,Aichernig2016}.
We will actually use a graph-based approach, whereby we roughly interpret Mealy machines 
as \glspl*{LTS} and build a synchronous product with respect to a conformance relation~\cite{Fernandez91,Weiglhofer2008}.
This will be described in more detail in Section~\ref{sec:case_study}. 


\paragraph{Comparison to Traditional Model-Based Testing}

Now that we introduced the approach, we can discuss the effort required to perform learning-based testing
in comparison to the effort required for traditional model-based testing. 
Some kind of adapter and abstraction component has to be implemented for both techniques, thus we address
the effort related to interpretation of requirements for a comparison.

Usually in model-based testing, a large set of requirements stated in natural language has to be formalised 
which is both labour-intensive and error-prone. This, however, is not required in the learning-based approach.
To perform the case study, we skimmed through the \gls*{MQTT}-specification to get a rough understanding
of the protocol in general. Afterwards, we identified interesting interactions 
between control packets and specific parameters thereof to implement mappers. This can be compared to the definition of
scenarios encoding test purposes, e.g , used by Spec Explorer~\cite{Grieskamp2006}. 
In addition to that, we only had to analyse parts of the specification in more depth which correspond to 
suspicious traces. 

Hence, less manual effort is required. It should be noted, though that this comes at the cost 
of decreased control of the testing process. While it is usually possible to direct the test case generation
through test selection criteria~\cite{utting2012}, the tests performed for learning are selected
by learning algorithms.

\paragraph{The Role of Learning}

A question that comes to mind concerns the role of learning in our approach. Why do we actually learn
Mealy machines? It would also be possible to generate some test suite, e.g. randomly, run the test suite
on all implementations and check whether differences in outputs exist. However, learning offers two benefits.
Firstly, it provides us with a model which can be used for further verification tasks such as model-checking~\cite{Fiterau_Brostean2016}.
In addition to that, it essentially defines a stopping criterion. Stated differently, we assume that we have
tested adequately and can stop testing as soon as we can derive a correct system model. A similar 
approach is, e.g., followed by Meinke and Sindhu~\cite{meinke2011}. They stop testing when learning converges,
but they also use other stopping criteria if learning does not converge. 

%% file: case_study.tex
\section{Case Study}

\label{sec:case_study}
In the following, we will discuss some implementation-specific details concerning both learning and conformance
checking of learned models. 
\subsection{Learning}

The learning part was implemented in Scala using the Java-library LearnLib for active automata learning~\cite{isberner2015}.
Most of the learning-related functionality was thus already implemented and we only had to implement 
application-specific components such as mappers, and a component responsible for the configuration of learning experiments. 

\subsubsection{Application-Specific Components}

We had to implement the three components shown in the middle of \figurename{} \ref{fig:learning_architecture},
i.e. the adapter, the client interface, and the mapper. While adapter and client interface merely perform parsing, serialisation,
and sending of packets, mappers specify a learning target. 

As noted in Section~\ref{sec:approach}, 
we used several sets of packet types with 
interesting dependencies within these sets as a basis for learning and consequently testing. 
For this purpose, we implemented seven different mappers,
which all use the same client interface and thereby can be used interchangeably to test different aspects.
Due to space limitations, we will describe only two of the seven implemented mappers.

\begin{compactdesc}
 \item [Simple:] The mapper \emph{Simple} controls one client and offers seven inputs exercising only
 basic functionality such as the simplest forms of subscribing and publishing.
 \item [Two Clients with Retained Will:]
 This mapper controls two clients, one of which sets a \emph{will message} while connecting and which may 
 close the \gls*{TCP} connection without properly disconnecting. The other client may subscribe to the topic 
 to which the will message is published. More specifically, the will message is published as retained message 
 which means that it is kept in the broker's state and sent whenever a client subscribes to the corresponding topic. 
\end{compactdesc}

It is of course possible to define further mappers for learning other functionality. However, we do not aim at completely testing 
\gls*{MQTT} brokers. We rather aim at showing that our approach is an effective aid at finding errors and we assume
that experiments performed with seven different mappers provide sufficient evidence for this purpose.

\subsubsection{Configuration}
\label{sec:configuration}

In order to evaluate different learning algorithms, equivalence checking algorithms and 
\gls*{MQTT} implementations, we needed to implement a configuration component
with which we can setup learning. However, a thorough comparison of algorithms is beyond the scope of
the paper. 

We used the TTT learning algorithm~\cite{isberner2013} for all experiments presented in the following
as it performed best in our experiments. Furthermore, 
we used the random-walk equivalence oracle provided by LearnLib to perform equivalence queries. This equivalence
oracle basically performs random tests to check whether the current hypothesis is correct. Although random testing
is not well-suited to guarantee coverage of some kind it is a valid choice in our context. The main reason for this
is the lack of scalability of more thorough methods like the W-method \cite{Chow1978,Vasilevskii1973} or the Wp-method~\cite{Fujiwara1991}.
These methods are computationally expensive in terms of runtime, even for only moderately complex models. In order to apply
those, it would be necessary to use low depth values (difference in number of states between hypothesis and actual models)
which limits their capability to find counterexamples. As a result, random testing was found to be better suited. 

We used random walks with the following settings:
\begin{compactitem}
 \item probability of resetting the \gls*{SUL}: $0.05$
 \item maximum number of steps: $10000$
 \item reset step count after finding a counterexample: $\mathit{true}$
\end{compactitem}


Additionally, there is a configuration parameter specifying an application-specific timeout on the
receipt of messages.

\subsection{Conformance Checking}
\label{sec:conf_check}
In the following, we will describe our approach to checking conformance between two learned models.
We actually check for equivalence and either output that the models are equivalent or present
all found counterexamples to the user. This is accomplished through the application of bisimulation checks.
%
 \paragraph*{"On the Fly"-Check}
 We implemented this equivalence checking method in a way similar to the bisimulation check in \cite{Fernandez91}.
 For this purpose, we interpreted Mealy machines as \glspl*{LTS} whereby we interpreted input-output pairs 
 labelling a transition in a Mealy machine as a single transition label in the \gls*{LTS}-interpretation. 
 
 In order to find counterexamples to equivalence we have to find \emph{fail}-states in a product graph 
 of the two considered models, created with respect to bisimulation. The graph contains states formed by pairs of states of both 
 Mealy machines, additional fail-states, and transitions between those states. A transition between ordinary
 states is added for input-output pairs executable in both Mealy machines and a transition to a fail-state is 
 added if a transition for some input-output pair is executable in only one of the Mealy machines, but not in the other. 
 We thereby check for observation equivalence because we add a fail-state only if there is some input for which the two Mealy
machines produce different outputs. 
 
 Since we consider deterministic Mealy machines, this check is simple to implement~\cite{Fernandez91}. We implemented it
 via an implicit depth-first search for fail-states in the product graph. 
 During this exploration, we collect all traces leading to fail-states and present them as counterexamples
 to the user. Since counterexamples are the only relevant information,
 we do not actually create the graph. It suffices to store visited states in order to avoid 
 exploring some state twice. 
 
 Note that the straight-forward implementation may also miss some bugs. 
 Consider a bug which merely produces wrong outputs but does not affect state transitions. 
 In this case the bisimulation check will stop exploring at the wrong output and add the reached state
 to the visited states. If, however, it is necessary to explore the model beyond this state to find another
 bug, we may miss this bug. Hence, it is possible to miss \emph{double faults} if both faults are reached by a single 
 trace. 
 
 To circumvent this problem, we added the possibility to extend counterexamples further until either another difference
 is found or a visited state is reached. Actually, the exploration can be continued until 
 a preset maximum number of differences along a trace has been found. 
 With this extended exploration it is thus possible to find
 multiple counterexamples in cases where the standard exploration would have found only one. As a result, the 
 effort required to analyse counterexamples is increased, but it may pay off if additional bugs are found. 
 
 The extended exploration did not uncover further bugs in our experiments, but led to a modification 
 of the learning setup. Two models were incorrectly learned because of insufficient equivalence
 testing (equivalence query). This issue was detected by cross-checking with models of other implementations with extended
 exploration. Consequently, the number of steps for random equivalence testing has been increased.

\subsection{Experiments}
In the following, we will discuss our case study in more detail. We will start by discussing the 
basic setup. 
Afterwards we will describe some of the bugs and differences between models
we found. In this context, we will consider difficulties and issues we faced as well as the 
manual effort required to classify counterexamples as failures. 

\subsubsection*{Setup}
\label{sec:setup}

We learned models of five freely available implementations of \gls*{MQTT} brokers, all of which are in active development
at the time of writing this paper. The brokers are (included in):
\begin{compactitem}
 \item Apache ActiveMQ 5.13.3\footnote{\url{http://activemq.apache.org/}}
 \item emqttd 1.0.2 \footnote{\url{http://emqtt.io/}}
 \item HBMQTT 0.7.1 \footnote{\url{https://github.com/beerfactory/hbmqtt}}
 \item Mosquitto 1.4.9 \footnote{\url{http://mosquitto.org/}}
 \item VerneMQ 0.12.5p4 \footnote{\url{https://vernemq.com/}}
\end{compactitem}

Since all brokers implement version 3.1.1 of \gls*{MQTT}, it was possible to perform all learning experiments
in the same way with only minor adaptations. The adaptations basically amount to specifying application-specific
timeouts for receiving packets. 
Table \ref{tab:timeout} shows the timeout values
used for the different implementations. We found these values via experiments and note that they are neither 
optimal nor do large timeout values indicate poor performance in general. A broker requiring a large timeout
may, e.g., provide excellent scalability to large numbers of connections which we did not test. 

\begin{table}[t]
\caption{Timeout values for receiving messages.}
\label{tab:timeout}
\centering
\scriptsize
 \begin{tabular}{|c|c|}
 \hline
  Implementation & Timeout in Milliseconds \\
  \hline 
  \hline 
  Apache ActiveMQ & 300 \\\hline
  emqttd & 25 \\\hline
  HBMQTT & 100 \\\hline
  Mosquitto & 100 \\\hline
  VerneMQ & 300\\\hline   
 \end{tabular}
\end{table}

All experiments were performed with a Lenovo Thinkpad T450 with 16 GB RAM and an Intel 
Core i7-5600U CPU operating at $2.6$ GHz and running Xubuntu Linux 14.04.

\subsection{Bug Hunt}
\label{sec:bug_hunt}

In the following, we will discuss our findings with respect to error detection in implementations.
Altogether we found $17$ bugs in all implementations combined whereby we did not find any in the Mosquitto
broker. Additionally, we found two cases of unexpected non-determinism in two implementations which hindered learning
and therefore are not included in the $17$ bugs found. Finally, we found a part of the specification
which strictly speaking none of the implementations implemented correctly. However, four of the implementations
showed behaviour users would expect to see thus we consider these implementations to be correct. 
The last implementation on the other hand showed faulty behaviour. Hence, \textbf{we actually found $\mathbf{18}$ bugs}. 

Four of the bugs correspond to issues already reported by other users. The remaining 
bugs were reported by us and are currently being reviewed or are already fixed by developers of the brokers.
We will give some examples showing bugs we found and highlighting issues we faced.

\subsubsection{Violations of the Specification}

\begin{figure}[t]
\centering
    \includegraphics[width=0.45\textwidth]{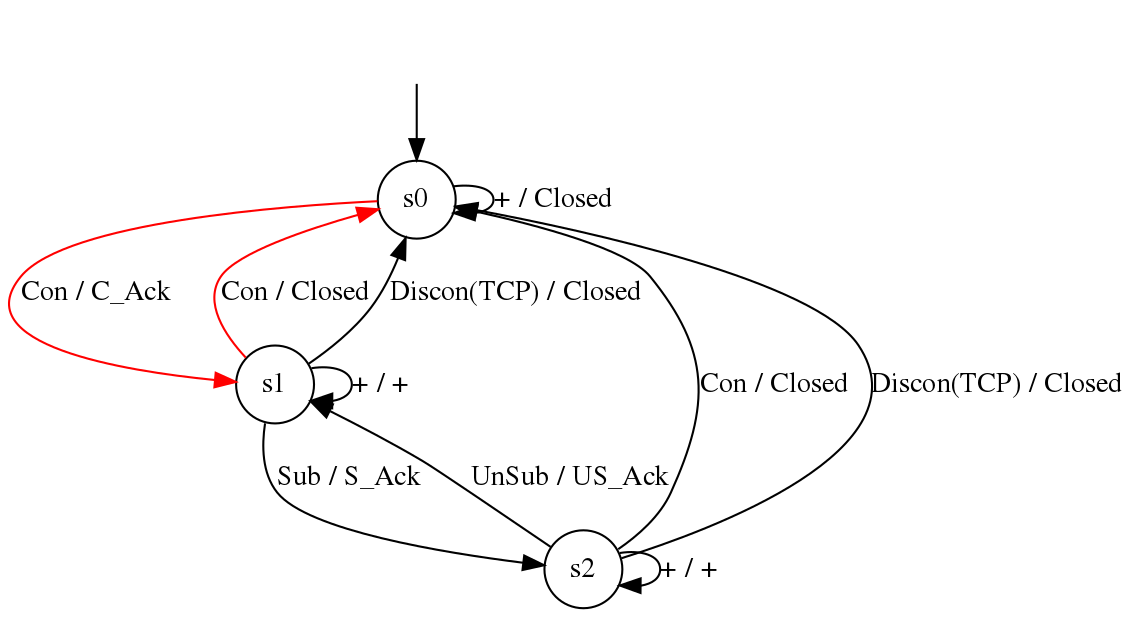}\\
    {(a) Mosquitto model}
    \includegraphics[width=.35\textwidth]{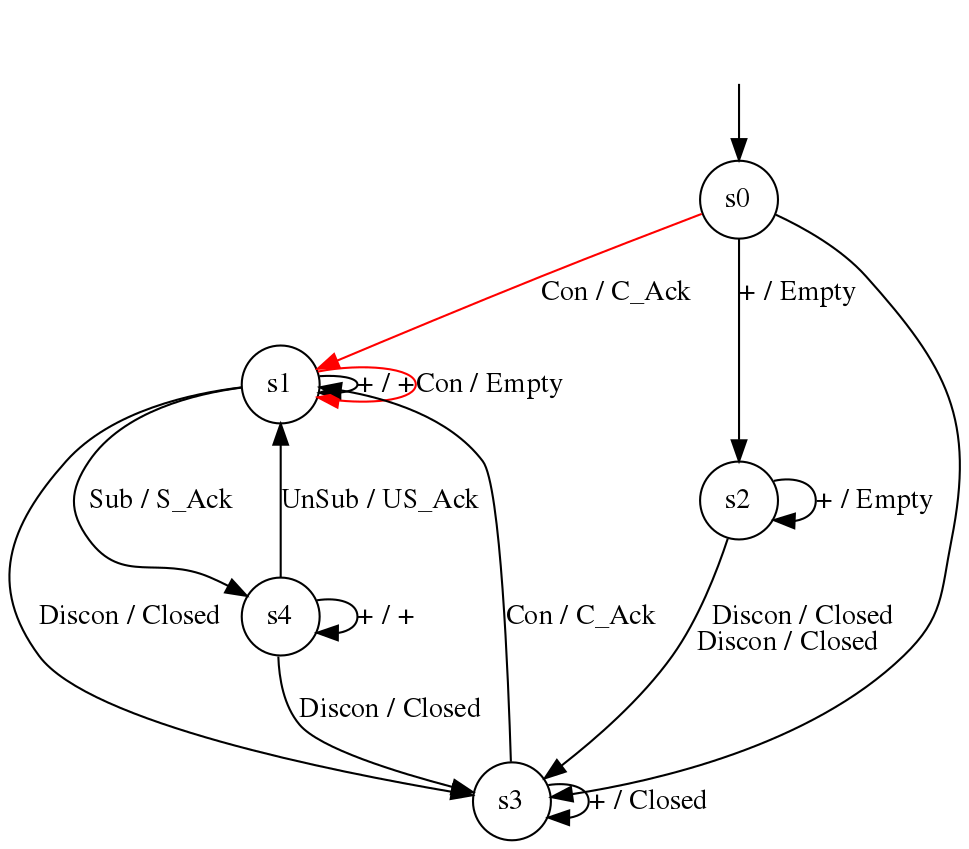}\\
    {(b) HBMQTT model}
   \caption{Models of two implementations learned with the \emph{Simple} mapper, whereby some inputs and outputs have been combined (denoted by +).}
   \label{fig:simple_mapper_models}
\end{figure}

A simple example of a violation of the protocol specification can be found by considering the behaviour
of the HBMQTT broker with respect to the functionality covered by the \emph{Simple} mapper. 
\figurename{} \ref{fig:simple_mapper_models} shows the models learned by observing the Mosquitto broker and the HBMQTT 
broker with abbreviated action labels. A counterexample to equivalence is $\var{Connect} \cdot \var{Connect}$,
which is shown in red in both models. For Mosquitto we have the output $\var{C\_Ack} \cdot \var{ConnectionClosed}$ and for
HBMQTT we have the output $\var{C\_Ack} \cdot \var{Empty}$, i.e. HBMQTT acknowledges the first connection attempt 
and ignores the second by not producing any output and it actually does not change its state as well.

The \gls*{MQTT} specification states that Mosquitto's behaviour is correct whereas HBMQTT behaves
in an incorrect way~\cite{MQTT_spec}:
\begin{quote}
 A Client can only send the CONNECT Packet once over a Network Connection. The Server MUST
process a second CONNECT Packet sent from a Client as a protocol violation and disconnect the Client
[MQTT-3.1.0-2].
\end{quote}

This is admittedly a rather simple example. However, we found also more subtle bugs. 
A strength with regard to error detection of our approach is that Mealy machines are
input enabled. Therefore, we do not only learn and test the usual behaviour, but also 
exceptional cases. Using the mapper \emph{Two Clients with Retained Will} 
we found an interesting sequence which uncovered bugs in both
emqttd and ActiveMQ.

The sequence is as follows:
\begin{compactenum}
 \item A client connects with client identifier \verb+Client1+ 
 \item A client connects with client identifier \verb+Client2+ with retained will message \verb+bye+ for topic \verb+c2_will+
 \item \verb+Client2+ disconnects unexpectedly (such that the will message is published)
 \item \verb+Client1+ subscribes to \verb+c2_will+
 \item \verb+Client1+ subscribes to \verb+c2_will+
\end{compactenum}
The responses to the first two steps are delivered as expected, which are acknowledgements and the will message
is sent to \verb+Client1+ in the fourth step which is also correct.
In the fifth step, Mosquitto behaves differently from emqtt and ActiveMQ. While emqttd and ActiveMQ do not resend \verb+bye+ to \verb+Client1+,
the Mosquitto broker sends \verb+bye+ again. 

The behaviour of ActiveMQ and emqtt is incorrect according to [MQTT-3.8.4-3] \cite{MQTT_spec} which states 
that repeated subscription requests must replace existing subscriptions and that ``any existing retained messages
matching the Topic Filter MUST be re-sent''.
\subsubsection{Non-Determinism}

An issue we faced during our experiments was non-deterministic behaviour with which Mealy-machine learning algorithms cannot
cope. In the case of our setup which is based on LearnLib, an exception is thrown and learning is stopped as soon as non-deterministic
behaviour is detected. Thus, we may waste test time in these cases. The only information we gain from such experiments is that
non-determinism affects the experiments accompanied with two input/output sequences with the same inputs but different 
outputs, i.e. sequences witnessing non-determinism. 

Non-determinism may result from several sources:
\begin{compactitem}
 \item learning setup
 \item time-dependent issues
 \item actual non-determinism displayed by implementations
\end{compactitem}
In the first case, it is actually beneficial that learning stops, as the setup should not introduce
non-determinism. It is likely to contain errors in this case. 
One issue related to time is the unknown time it takes for a broker to respond. In order to avoid 
non-deterministic behaviour in this regard, we implemented the aforementioned timeout on the receipt
of messages. 
%

We thus introduce imprecision to overcome time-related non-determinism. Considering 
that \gls*{TCP} is actually a reliable protocol and that the \acrfull*{UDP} is often used in the \gls*{IoT}, time-related
non-determinism is likely to be a more severe issue in other \gls*{IoT} protocols. 

Implementations may also show truly non-deterministic behaviour, i.e. the repeated execution of some input sequence 
under the same conditions may lead to different results. Unfortunately, we cannot adapt our learning setup 
with reasonable effort to account for actual non-determinism. For this reason, we could not 
successfully perform $3$ out of $35$ learning experiments for the bug hunt. 
 We actually evaluated further \gls*{MQTT} implementations in addition to those listed in Section~\ref{sec:case_study}, 
which we excluded from the experiments because of non-deterministic behaviour.
These additional implementations showed non-deterministic behaviour during learning with even the most simple mapper. 
Considering that, we conclude that 
learning-based verification would greatly benefit from being able to learn non-deterministic models. 


\subsubsection{Discussion}

In the following, we will review our experiences using the proposed approach with special
regard to manual effort. In this context, we will also recapitulate some already discussed 
issues affecting the required effort. 

In the initial phase it requires some experimentation to define mappers with reasonable complexity,
i.e.\ such that learning is possible in an acceptable amount of time. This usually
does not require a substantial amount of human labour, but requires computation time as experiments
have to be executed repeatedly with at least one implementation. This can 
probably best be compared to defining test-case specification scenarios, e.g., for testing with Spec Explorer~\cite{Grieskamp2006}. 

However, we also spent a significant amount of time analysing suspicious traces, 
a task not needed in traditional model-based testing as requirements have to be formalised beforehand.
In this context, we made the observation that bugs usually result in several counterexamples to 
equivalence.  In addition to the standard equivalence check, we also used the extended 
bisimulation check to avoid missing bugs. This, however, required additional
manual effort. 

Consider for instance the first bug discussed and highlighted in \figurename{}~\ref{fig:simple_mapper_models}.
Essentially the same bug can be detected by analysing the counterexample $\var{Connect} \cdot \var{Subscribe} \cdot \var{Connect}$.
Cross-checking the models in \figurename{} \ref{fig:simple_mapper_models}, 
we actually found $7$ counterexamples with the bisimulation check and $24$ with the bisimulation check performing extended exploration.
All these counterexamples point to only two different bugs. An example in which the extended check does not cause any overhead
is related to a bug of VerneMQ
which causes the broker to not publish empty retained messages. Checking equivalence between a model of Mosquitto 
and a model of VerneMQ learned with a mapper not described in this paper finds
$4$ counterexamples with either of the checks.

At the current stage, we implemented a mechanism to manually define filters to hide counterexamples
matching a specified pattern. Thus, it is possible to analyse a counterexample, find a bug 
and specify a pattern to exclude similar counterexamples. Coarse patterns may lead to bugs being undetected,
therefore we did not use filters in our experiments.

However, a reduction of counterexamples or some kind of automated partitioning into 
equivalence classes of counterexamples may be crucial for a successful application 
of the approach to more complex systems. 
To implement such a technique it would, e.g., be possible to follow an approach similar to \emph{MoreBugs}~\cite{DBLP:conf/icse/HughesNSA16}.
The MoreBugs method tries to infer a bug pattern from a failing test case and avoids testing the same pattern repeatedly.
We could group counterexamples by matching them to inferred patterns and present only one counterexample per group to users.
Especially parallel-composition-based pattern inference seems promising in our use case. Since
Mealy machines are input-enabled, inputs result in self-loops in many states which causes counterexample traces to be interleaved
with non-relevant inputs.

We noted in Section~\ref{sec:approach} 
that it may be possible to learn an incorrect 
model if the equivalence oracle which is only an approximation provides a wrong answer. Therefore,
we test counterexamples as stand-alone tests as well to see whether they are spurious. In this way, we actually create
a regression test suite focused on previously detected bugs. 

A more problematic scenario is that we may learn incompletely and thereby overlook erroneous behaviour.
However, on the one hand we have seen that bugs usually result in several counterexamples which lowers the probability
of missing bugs. On the other hand, testing is inherently incomplete, so there is always the possibility
that we do not detect all bugs. 

We conclude that it is possible to find non-trivial bugs in protocol implementations 
with reasonable effort despite necessary harsh abstraction. 
Testing more complex systems may, however, be hindered by the large number of counterexamples 
that need to be analysed. 
Tasks other than that have comparable counterparts in traditional model-based testing.
It should be emphasised that the initial effort to setup a learning environment 
is relatively low due to the flexibility and ease of use of LearnLib~\cite{isberner2015}.

\subsection{Efficiency}
\label{sec:efficiency}

We faced an issue during learning which is related to runtime. To illustrate 
the severity of this problem, Tables \ref{tab:bug_hunt_simple} and \ref{tab:two_client_will} 
show runtime measurement results for learning with the 
two described mappers. 

The results include the number of states in the learned models, 
the time and number of queries needed for membership queries (\emph{MQ time[s]} and \emph{MQ \# queries}), 
and the time and number of queries needed for conformance testing (\emph{CT time[s]} and \emph{CT \# queries}).
The number of queries for conformance testing reflects the actual number of tests carried
out to check equivalence. In other words, this number denotes the number of conformance tests
executed during the equivalence queries performed throughout learning.
The number of equivalence queries 
represents the number of rounds of every learning experiment, i.e. the last row shows the number
of hypotheses constructed by learning.

\begin{table}[t]
\scriptsize
\begin{center}
\caption{Experimental results obtained by learning with the mapper \emph{Simple}
with an alphabet size of $7$.}
\label{tab:bug_hunt_simple}
\begin{tabular}{|c|c|c|c|c|c|}\hline
                                    & ActiveMQ & emqttd   & HBMQTT   & Mosquitto & VerneMQ \\\hline
\# states                              & $4$ & $3$ & $5$ & $3$ & $3$ \\\hline
\specialcell[t]{MQ time[s]}       & $59.72$ & $3.87$ & $31.94$ & $14.01$ & $43.91$ \\\hline
\specialcell[t]{MQ  \# queries }  & $88$ & $59$ & $110$ & $56$ & $57$ \\\hline
\specialcell[t]{CT  time[s]}       & $914.18$ & $78.3$ & $491.06$ & $278.21$ & $915.77$  \\\hline
\specialcell[t]{CT  \# queries }  & $525$ & $519$ & $482$ & $487$ & $490$ \\\hline
\specialcell[t]{\# equivalence \\ queries }   & $4$ & $3$ & $4$ & $3$ & $3$ \\\hline
\end{tabular}
\end{center}
\end{table}

It can be observed that we actually deal with relatively simple models. 
The largest models have eighteen states and the larger of the two alphabets contains nine input symbols.
Despite the possibility to learn much larger models with active automata learning, e.g., Merten 
et al.\ noted that they achieved to learn a system with over a million states~\cite{merten2011}, 
we still faced efficiency issues. This can be explained by considering the long runtime of individual tests/queries.

In our setting, tests may take several seconds since we wait up to $600$ milliseconds for outputs from 
two clients in response to a single input ($300$ milliseconds per client). Thus, we see similar learning 
performance as when learning the \gls*{TLS} protocol~\cite{de_ruiter2015}. However, unlike in the context of learning
\gls*{TLS}, we do not stop testing once a connection is closed. Since there are persistent 
sessions and other related features of \gls*{MQTT} we also learn behaviour relevant to, e.g., session resumption. 

The drastic influence of testing runtime can be seen in experiments performed with ActiveMQ and VerneMQ as they require
the largest timeout value on the receipt of outputs. Even the simplest model of VerneMQ takes almost $16$ minutes to learn (see Table~\ref{tab:bug_hunt_simple}).
The longest experiment, learning a model of ActiveMQ with the 
mapper \emph{Two Clients Retained Will}, takes more than $110$ minutes and resulted in a model with only eighteen states 
(see Table~\ref{tab:two_client_will}). These high computation times for learning comparably simple models 
make apparent that there is a need to keep the number and length of queries to be executed as small as possible. This 
can, e.g., be achieved  
via domain-specific optimisations, heuristics and smart test selection~\cite{Hungar2003,Smeenk2015}, or via algorithmic advantages~\cite{isberner2013}.


\begin{table}[t]\scriptsize
\begin{center}
\caption{Experimental results obtained by learning with the mapper \emph{Two Clients with Retained Will}
with an alphabet size of~$9$.}
\label{tab:two_client_will}
\begin{tabular}{|c|c|c|c|c|c|}\hline
                                    & ActiveMQ & emqttd   & HBMQTT   & Mosquitto & VerneMQ \\\hline
\# states                              & $18$ & $18$ & $17$ & $18$ & $17$ \\\hline
\specialcell[t]{MQ  time[s]}       & $1855.55$ & $167.32$ & $557.14$ & $641.89$ & $1570.8$ \\\hline
\specialcell[t]{MQ  \# queries }  & $732$ & $735$ & $640$ & $730$ & $625$ \\\hline
\specialcell[t]{CT  time[s]}       & $4787.92$ & $481.36$ & $2022.47$ & $1612.59$ & $4355.97$  \\\hline
\specialcell[t]{CT  \# queries }  & $672$ & $816$ & $613$ & $670$ & $658$ \\\hline
\specialcell[t]{\# equivalence \\ queries }   & $13$ & $12$ & $11$ & $9$ & $11$ \\\hline

\end{tabular}
\end{center}
\end{table}

%% file: conclusion.tex
\section{Conclusion}
\label{sec:discussion}
\subsection{Summary}

In this paper we presented a learning-based approach to semi-automatically detect 
failures of reactive systems. We evaluated the effectiveness of this approach by means
of a case study. In total we found $18$ faults in four out of five \gls*{MQTT} brokers. 

More concretely, we learned abstract models of \gls*{MQTT} brokers. Based on that, we identified observable differences 
between the considered implementations in an automated manner. 
Since these differences are likely to show erroneous behaviour we
inspected manually whether they show specification violations.

To the best of our knowledge, we presented the first 
such case study focusing on reactive systems implemented independently by open-source developers
and it is the first attempt at model-based testing \gls*{MQTT} brokers.
We showed that 
the proposed approach can be effective at detecting bugs without requiring any prior modelling. 
Additionally, we showed that interactions requiring a long time to complete can be an obstacle. It is a known
fact that active automata learning shows efficiency problems while learning models with large input 
alphabets and state space~\cite{Berg2005a}. This issue is especially problematic when dealing 
with systems with long and unknown response delays, a property exhibited by \gls*{MQTT} implementations. 

While the approach can generally be applied to any type of reactive system for which there 
exist multiple implementations, it is especially well-suited to protocol testing because:
(1) protocols can be modelled abstractly with low numbers of states, making active learning feasible, and
(2) well-defined standards are likely to exist for common protocols. However, another possible
application scenario is regression testing of reactive systems~\cite{Hungar2003}, i.e. a model of a 
new system version could be learned and checked for conformance to the model of a previous version.

\subsection{Future Work}


As noted before, we had to deal with non-determinism. It
is actually common for complex reactive systems to behave non-deterministically so it may be worthwhile
to investigate ways to learn non-deterministic models of reactive systems such as 
non-deterministic Mealy machines~\cite{Khalili_Tacchella2014} or \glspl*{IOTS}~\cite{Volpato_Tretmans2015}.
Both of the cited approaches unfortunately suffer from the fact that we can never be sure when 
we have seen all non-deterministic behaviours. In other words, we do not know how often we need to apply
some sequence of inputs in order to see all possible responses. In practice, assumptions have to be made
with which we can derive bounds on the number of required repetitions of some input sequence. 
Through the requirement of repeated executions, however, the computation time increases
which is already an issue. 

Alternatively, output non-determinism~\cite{Khalili_Tacchella2014} may be resolved 
by learning probabilistic rather than non-deterministic models.
There exist promising passive learning approaches, e.g., based on state merging,
which infer (probabilistic) models from samples obtained prior to learning~\cite{Carrasco_oncina_1994,delaHiguera_2010}.
Such methods have already been investigated in a verification context~\cite{Mao_2016}.

We also observed that there exists time-dependent behaviour and that such behaviour is likely to play
a more crucial role in other \gls*{IoT} protocols using means of communication less reliable than \gls*{TCP}. 
As a result, there is a need for an investigation of appropriate methods to infer models of timed systems in this area.
There exist approaches for this task, in an active setting for learning event-recording automata \cite{Grinchtein2006,Grinchtein2010,wei_lin_2011}
and in a passive setting for learning timed automata with a single clock~\cite{Verwer2011,Verwer2012}. 
Both approaches place restrictions on the expressiveness of models they consider.
The former, however, is less restricted at the expense of higher worst-case complexity. Hence, it would be interesting
to examine practical limits of these approaches and whether they could be improved in terms 
of performance or learnability. Verwer et al.~\cite{Verwer2011} identified two intriguing starting points
for future research concerning learnability of more expressive timed automata: (1) they showed that in general timed automata
with multiple clocks cannot be efficiently identified. However, specific classes of timed automata 
might be efficiently identifiable. (2) Timed automata with $n$ clocks can actually be
represented by the intersection of $n$ timed automata with only one clock. Hence, a possible approach
would be to learn multiple one-clock timed automata in a first step and combine them in a second step. 

A possible extension to the \gls*{MQTT} case study would be to infer 
models of client implementations as well and to verify properties of the composition 
of clients and brokers. Such an approach has been followed by Fiter{\u{a}}u-Bro{\c{s}}tean et al.
for analysing \gls*{TCP} implementations~\cite{Fiterau_Brostean2016}.